\documentclass[aip,cha,12pt,floatfix]{revtex4-1}

\usepackage{graphics}
\usepackage{graphicx}

\usepackage{color}

\begin{document}


\title{Complex Statistics and Diffusion in Nonlinear Disordered Particle Chains} 



\author{Ch. G. Antonopoulos}
\email[]{chris.antonopoulos@abdn.ac.uk}
\affiliation{Institute for Complex Systems and Mathematical Biology (ICSMB), Department of Physics, University of Aberdeen, AB24 3UE Aberdeen, United Kingdom}

\author{T. Bountis}
\email[]{bountis@math.upatras.gr}
\affiliation{Center for Research and Applications of Nonlinear Systems (CRANS), Department of Mathematics, University of Patras, 26500 Patras, Greece}

\author{Ch. Skokos}
\email[]{haris.skokos@uct.ac.za}
\affiliation{Department of Mathematics and Applied Mathematics, University of Cape Town, Rondebosch, 7701, South Africa}
\affiliation{Department of Physics, Aristotle University of Thessaloniki, 54124 Thessaloniki, Greece}

\author{L. Drossos}
\email[]{ldrossos@teimes.gr}

\affiliation{High Performance Computing Systems Lab (HPCS lab), Department of Computer and Informatics Engineering, Technological Educational Institute of Western Greece, 30300 Antirion, Greece}




\date{\today}

\begin{abstract}
We investigate dynamically and statistically diffusive motion in a Klein-Gordon particle chain in the presence of disorder. In particular, we examine a low energy (subdiffusive) and a higher energy (self-trapping) case and verify that subdiffusive spreading is always observed. We then carry out a statistical analysis of the motion in both cases in the sense of the Central Limit Theorem and present evidence of different chaos behaviors, for various groups of particles. Integrating the equations of motion for times as long as $10^9$, our probability distribution functions always tend to Gaussians and show that the dynamics does {\em not} relax onto a quasi-periodic KAM torus and that diffusion continues to spread chaotically for arbitrarily long times. 

\end{abstract}


\maketitle 


\textbf{
The absence of diffusion in disordered media, often called Anderson localization, is a general phenomenon that applies to the transport of different types of classical or quantum waves. An interesting question is what happens to the diffusion if nonlinearity is introduced. Many studies so far have focused on the evolution of an initially localized wave packet showing that it spreads subdiffusively for moderate nonlinearities, while for stronger ones a substantial part of it remains self-trapped. Currently, a greatly debatable problem concerns the long time spreading of the wave packet. It has been conjectured that chaotically spreading wave packets will asymptotically approach Kolmogorov-Arnold-Moser torus-like structures in phase-space, while numerical simulations typically do not show any sign of slowing down of the spreading behavior. Here, we introduce the concept of $q$-exponential statistics to shed new light on this problem. Thus, in the case of a high-dimensional Klein-Gordon disordered particle chain we concentrate on a low energy (subdiffusive) and a higher energy (self-trapping) case and verify that subdiffusive spreading always occurs following specific power-laws. Integrating the equations of motion for long times and computing probability distributions of sums of the positions of particles, we find convincing evidence that the dynamics does not relax onto a quasi-periodic Kolmogorov-Arnold-Moser torus-like structure, but continues to spread chaotically along the Klein-Gordon chain of particles for arbitrarily long times.}

\section{Introduction}\label{intro}

Probability distribution functions (pdfs) of chaotic trajectories of dynamical systems have been studied for many decades and by many authors, aiming to understand the transition from deterministic to stochastic dynamics \cite{Arnold1967,Sinai1972,Eckmann1985}. One of the most relevant and fundamental questions concerns the existence of an appropriate invariant probability density (or ergodic measure), characterizing chaotic motion in phase space regions where solutions generically exhibit exponential divergence of nearby trajectories. If it is possible to define such an invariant measure for almost all initial conditions (i.e. except for a set of measure zero), then one has a firm basis for studying the system from a statistical mechanics point of view.

Now, if this invariant measure is a continuous and sufficiently smooth function of the phase-space coordinates, one can invoke the Boltzmann-Gibbs microcanonical ensemble and attempt to evaluate all relevant quantities of equilibrium statistical mechanics, like partition function, free energy, entropy, etc. of the system under study. On the other hand, if the measure is absolutely continuous (as e.g. in the case of the so-called Axiom A dynamical systems), one might still be able to use the formalism of modern ergodic theory to study the statistical properties of the model \cite{Eckmann1985}.

Since the existence of an invariant measure is not known a priori, one may still proceed in the context of the Central Limit Theorem\cite{Rice1995} (CLT) and consider the values of one (or a linear combination) of components of a chaotic solution at discrete times $t_i,\;i=1,\ldots,M$ as realizations of $\mathcal{N}$ independent and identically distributed (iid) random variables $X^{(j)}(t_i),\;j=1,\ldots,\mathcal{N}$. If the motion under study is uniformly chaotic (ergodic) in some region of phase space, one typically finds that the pdfs of the {\em sums} of these variables converge rapidly to a Gaussian distribution, whose mean and variance are those of the $X^{(j)}$'s. In such cases (which we call ``strongly'' chaotic) at least one Lyapunov exponent is positive and the respective subset of the constant energy manifold is uniformly covered by chaotic orbits, for all but a (Lebesgue) measure zero set of initial conditions.

What happens, however, if the motion is not uniformly chaotic and the orbits ``stick'' for long times on the boundaries of islands surrounding stable periodic orbits, where Lyapunov exponents become very small and may even vanish? In such regimes, the motion is often termed ``weakly'' chaotic, as trajectories get trapped within complicated sets of cantori and diffuse slowly through multiply connected domains in a highly non-uniform way \cite{Aizawa1984,Chirikov1984,Meiss1986}. Many such examples occurring in physically realistic systems have been studied in the recent literature (see for example Refs. \cite{Skokos2008,Flach2009,Johansson2009,Skokos2009}).

In this paper, we investigate the existence of possible connections between such regimes of ``weak'' and ``strong'' chaos and subdiffusive motion in the presence of disorder by considering a Hamiltonian particle chain in the presence of nonlinearity and disorder. In particular, we demonstrate first that pdfs of sums of position variables in this system \textit{do not} rapidly converge to a Gaussian distribution, but are well approximated for long times by the so-called $q$-Gaussian distribution \cite{Tsallisbook2009}
\begin{equation}\label{q_gaussian}
\mathtt{P}(s)=a \exp_q({-\beta s^2})\equiv a\biggl[1-(1-q)\beta s^{2}\biggr]^{\frac{1}{1-q}}
\end{equation}
where the $q$ entropic index satisfies $1<q<3$, $\beta$ is an arbitrary parameter and $a$ is a normalization constant. Eventually, of course, chaotic orbits seep out from smaller regions to larger chaotic seas, where obstruction by islands and cantori is less dominant and the dynamics is more uniformly chaotic. This transition is signalled by the $q$ entropic index of the distribution (\ref{q_gaussian}) decreasing towards $q=1$, which represents the limit at which the pdf becomes a Gaussian distribution.

Thus, we shall speak of ``weak'' chaos when the value of the entropic index $q$ is greater than unity by at least one decimal point, e.g. $q=1.1$ or higher (with $q<3$) and the corresponding pdfs are distinctly different from a Gaussian pdf. On the other hand, if $q$ is closer to 1 we speak of ``strong'' chaos, where the associated pdfs become practically indistinguishable from a Gaussian pdf, see e.g. our Fig. \ref{distributions_N1000_E0.4_different_etas} in the text, where panel a) depicts a strongly chaotic and panel b) a weakly chaotic case, respectively.

Concerning subdiffusion in a Hamiltonian system representing a disordered Klein-Gordon (KG) chain of $N=1000$ particles \cite{BLSKF11,Laptyeva_crossover_2010}, we find that even though there are intervals of weak chaos, strongly chaotic dynamics eventually prevails characterized by $q$-Gaussian pdfs that always approach a Gaussian pdf for long enough times. Thus, we suggest that the motion of this system will never approach a KAM regime of invariant tori as suggested by some authors \cite{Johansson_kam_2010,A11}.

This paper is organized as follows: In Sec. \ref{CLT_approach} we outline the details of our study of the statistical distributions corresponding to weakly and strongly chaotic behavior and Sec. \ref {KG_results} describes diffusive motion in a disordered Klein-Gordon chain. Finally, Sec. \ref{conclusions} contains our conclusions.

\section{Computation of Statistical Distributions of Weak and Strong Chaos}\label{CLT_approach}

In this work, we investigate the statistical properties of chaotic diffusion in a disordered Klein-Gordon chain. It is described by an autonomous $N$ degree of freedom Hamiltonian of the form
\begin{equation}
H\equiv H(x(t),p(t))=H(x_1(t),\ldots,x_N(t),p_1(t),\ldots,p_N(t))=E,\label{Ham_fun}
\end{equation}
where $(x_l(t), p_l(t)),\;l=1,\ldots,N$ are the positions and momenta respectively of the system in continuous time $t$. As is well-known, the solutions (or orbits) can be periodic, quasi-periodic or chaotic depending on the initial conditions and the values of their parameters. What we wish to explore here is the statistics of their diffusive dynamics in regimes of weakly chaotic motion, where Lyapunov exponents are positive but very small. Such situations often arise when solutions move slowly through thin chaotic layers, wandering through a complicated network of higher order resonances, often sticking for very long times to the boundaries of islands constituting the so-called ``edge of chaos'' regime \cite{Tsallisbook2009}.

Many interesting questions can be asked in this context: How long do these weakly chaotic states last? What type of pdfs characterize them and how can one connect them to the diffusion properties of the system? Does disorder in the choice of their parameter values play a role in these considerations? 

To answer such questions, we use the solutions of the equations of motion of our Hamiltonian system
to construct pdfs of suitably rescaled sums of $M$ values of a generic observable $\eta_i=\eta(t_i),\;i=1,\ldots,M$ which depends linearly on the position coordinates of the solution. Viewing these as iid random variables (in the limit of $M\rightarrow\infty$), we evaluate their sum
\begin{equation}\label{sums_CLT}
S_M^{(j)}=\sum_{i=1}^M\eta_i^{(j)}
\end{equation}
for $j=1,\ldots,N_{\mbox{ic}}$ different initial conditions and study the statistics of Eq. (\ref{sums_CLT}) centered about their mean value $\langle S_M^{(j)}\rangle=\frac{1}{N_{\mbox{ic}}}\sum_{j=1}^{N_{\mbox{ic}}}\sum_{i=1}^{M}\eta_i^{(j)}$ and rescaled by their standard deviation $\sigma_M$
\begin{equation}
s_M^{(j)}\equiv\frac{1}{\sigma_M}\Bigl(S_M^{(j)}-\langle S_M^{(j)}\rangle \Bigr)=\frac{1}{\sigma_M}\Biggl(\sum_{i=1}^M\eta_i^{(j)}-\frac{1}{N_{\mbox{ic}}}\sum_{j=1}^{N_{\mbox{ic}}}\sum_{i=1}^{M}\eta_i^{(j)}\Biggl),
\end{equation}
where
\begin{equation}
\sigma_M^2=\frac{1}{N_{\mbox{ic}}}\sum_{j=1}^{N_{\mbox{ic}}}\Bigl(S_M^{(j)}-\langle S_M^{(j)}\rangle \Bigr)^2=\langle S_M^{(j)2}\rangle -\langle S_M^{(j)}\rangle^2.
\end{equation}
Plotting the normalized histogram of the probabilities $\mathtt{P}(s_M^{(j)})$ as a function of $s_M^{(j)}$, we then compare our pdfs with a $q$-Gaussian of the form
\begin{equation}
\mathtt{P}(s_M^{(j)})=a\exp_q({-\beta s_M^{(j)2}})\equiv a\biggl[1-(1-q)\beta s_M^{(j)2}\biggr]^{\frac{1}{1-q}},
\end{equation}
cf. (\ref{q_gaussian}), where $q$ is the so-called entropic index. Note that this is a generalization of the well-known Gaussian pdf, since in the limit $q\rightarrow 1$ we have $\exp_q(-\beta x^2)\rightarrow\exp(-\beta x^2)$. Moreover, it can be shown that the $q$-Gaussian distribution (\ref{q_gaussian}) is normalized when
\begin{equation}\label{beta-$q$-Gaussian}
\beta=a\sqrt{\pi}\frac{\Gamma\Bigl(\frac{3-q}{2(q-1)}\Bigr)}{(q-1)^{\frac{1}{2}}\Gamma\Bigl(\frac{1}{q-1}\Bigr)},
\end{equation}
where $\Gamma$ is the Euler $\Gamma$ function. Clearly, Eq. (\ref{beta-$q$-Gaussian}) shows that the allowed values of $q$ are $1<q<3$ for this normalization.

The index $q$ appearing in Eq. (\ref{q_gaussian}) is connected with the Tsallis entropy \cite{Tsallisbook2009}
\begin{equation}\label{Tsallis entropy}
S_q=k\frac{1-\sum_{i=1}^W \mathcal{P}_i^q}{q-1}\mbox{ with }\sum_{i=1}^W \mathcal{P}_i=1,
\end{equation}
where $i=1,\ldots,W$ counts the microstates of the system, each occurring with a probability $\mathcal{P}_i$ and $k$ is the well-known Boltzmann constant. Just as the Gaussian distribution represents an extremal of the Boltzmann-Gibbs entropy $S_{\mbox{BG}}\equiv S_1=k\sum_{i=1}^W \mathcal{P}_i\ln \mathcal{P}_i$, so is the $q$-Gaussian (\ref{q_gaussian}) derived by optimizing the Tsallis entropy of Eq. (\ref{Tsallis entropy}) under appropriate constraints.

Systems characterized by the Tsallis entropy are said to lie at the ``edge of chaos'' and are significantly different than Boltzmann-Gibbs systems, in the sense that their entropy is nonadditive and generally nonextensive \cite{Tsallisbook2009}. In fact, a $q$-Central Limit Theorem has been proved \cite{UmarovTsallis2008} for $q$-Gaussian distributions (\ref{q_gaussian}) that is of the same form as the classical CLT.

Let us now describe the numerical aspects of the calculation of the above pdfs. First of all, in every case under study, we specify an observable denoted by $\eta(t)$ as one (or a linear combination) of the position variables of a chaotic solution.

Then, we divide the time interval of the evolution of the orbit into a predefined fixed number of $N_{\mbox{ic}}$ equally spaced, consecutive time windows, which are long enough to contain a significant part of the orbit. Next, we subdivide each of the $N_{\mbox{ic}}$ time windows into a fixed number $M$ of equally spaced subintervals and calculate the sum $S_M^{(j)}$ of the values of the observable $\eta(t)$ at the \textit{right edges} of these subintervals (see Eq. (\ref{sums_CLT})). In this way, we treat the point at the beginning of every time window as a new initial condition and repeat this process $N_{\mbox{ic}}$ times to obtain as many sums as required for reliable statistics. 

As we shall see in the next sections, in regions of weak chaos these distributions are well-fitted by a $q$-Gaussian for fairly long evolution intervals, whose $q$ value is distinctly greater than 1. It may happen, of course, for longer times that the orbits begin to diffuse through domains of strong chaos, in which case $q$ tends to 1 and the well-known form of a Gaussian pdf is recovered.

\section{Diffusive Dynamics of the Disordered Klein-Gordon Chain}\label{KG_results}
\subsection{The Disordered Quartic Klein-Gordon Model}\label{KG_model}

The absence of diffusion in disordered media (the so-called \emph{Anderson localization}\cite{Ander}) is a general phenomenon that applies to the transport of different types of classical or quantum waves, like electromagnetic, acoustic and spin waves. It is interesting to ask what happens if nonlinearity is introduced to the disordered system. Understanding the effect of nonlinearity on the localization properties of wave packets in disordered systems has attracted the attention of many researchers to date \cite{pikovsky_destruction_2008,kopidakis_absence_2008,Flach2009,Skokos2009,veksler_spreading_2009,SF10,Flach_spreading_2010, Laptyeva_crossover_2010,VKF10,MP10,MAP11,BLSKF11,BLGKSF11,BouSkobook2012}. Most of these studies consider the evolution of an initially localized wave packet and show that it spreads subdiffusively for moderate nonlinearities, while for strong enough nonlinearities a substantial part of it is self-trapped. In such works, one typically analyzes normalized norm or energy distributions $z_{l}\equiv E_{l}/\sum_{i=1}^{N} E_{i} \geq 0,\;l=1,\ldots,N$ and measures the second moment
\begin{equation}\label{second_moment}
	m_2= \sum_{l=1}^{N} (l-\bar{l})^2 z_{l},
\end{equation}
where $\bar{{l}} = \sum_{l=1}^{N} l z_{l}$, which is an efficient measure of the wave packet's spreading. In particular, for single-site excitations the wave packet's spreading leads to an increase of the second moment according to $m_2 \sim t^{1/3}$, both in the diffusive as well as the self-trapping case\cite{pikovsky_destruction_2008,Flach2009,Skokos2009,veksler_spreading_2009}.

Currently, a greatly debatable problem is the long time behavior of wave packet spreading in disordered nonlinear lattices. Recently it was conjectured\cite{Johansson_kam_2010,A11} that chaotically spreading wave packets will asymptotically approach Kolmogorov-Arnold-Moser (KAM) torus-like structures in phase-space, while numerical simulations typically do not show any sign of slowing down of the spreading behavior\cite{Laptyeva_crossover_2010,BLSKF11,SGF13}. Nevertheless, for particular disordered nonlinear models some numerical indications of a possible slowing down of spreading have been reported \cite{PF11,MP13}. Thus, we decided to implement the ideas of Tsallis statistics to shed new light on this problem.

For this purpose, we consider the quartic Klein-Gordon lattice described by the Hamiltonian of $N$ degrees of freedom
\begin{equation}
H_{KG}= \sum_{l=1}^{N} \frac{p_{l}^2}{2} + \frac{\tilde{\epsilon}_{l}}{2}
x_{l}^2 + \frac{1}{4} x_{l}^{4}+\frac{1}{2W}(x_{l+1}-x_l)^2=E,
\label{RQKG}
\end{equation}
where $x_l$ and $p_l$ are respectively the generalized positions and momenta on site $l$, and $\tilde{\epsilon}_{l}$ are chosen uniformly randomly from the interval $\left[\frac{1}{2},\frac{3}{2}\right]$ to account for the disorder present at each site $l$. This Hamiltonian conserves the value of the total energy $E\geq 0$ of the system, which, for fixed disorder strength $W$, serves as a control parameter of the nonlinearity. In our study, we follow the evolution of single site excitations by solving the equations of motion
\begin{equation}		
\ddot{x}_{l} = - \tilde{\epsilon}_{l}x_{l} -x_{l}^{3} + \frac{1}{W}
(x_{l+1}+x_{l-1}-2x_l),\;l=1,\ldots,N,
\label{KG-EOM}
\end{equation}
and monitor normalized energy density distributions.

In the next section, we first present the results of our numerical experiments describing the chaotic dynamics of wave packets in a KG chain of $N=1000$ particles for the low energy (subdiffusive) as well the high energy (self-trapping) case. We then carry out an analysis of the {\em statistics} of the motion in the sense of the CLT and find, in both cases, convincing evidence of initially weak and eventually strong chaos, for times as long as $10^9$! Indeed, our results show {\em no sign} of quasi-periodic KAM behavior and serve to further strengthen the conjecture that waves spread subdiffusively and chaotically for arbitrarily long times in nonlinear disordered media.

We use two representative examples of energies, $E=0.4$ (subdiffusive spreading) and $E=1.5$ (self-trapping) as reported in the work of Skokos et al.\cite{Skokos2009} and integrate numerically the KG chain using a fourth order Yoshida's symplectic integrator \cite{Yoshida1990}, which is very efficient for long integrations (e.g. up to $10^{9}$ time units) of lattices having typically $N=1000$ sites, to keep the required computational time at feasible levels, preserving at the same time the energy of the system to satisfactory accuracy. In particular, an integration time step $\tau=0.05$ typically keeps the relative energy error at about $10^{-6}$. In both cases, we consider one disorder realization, i.e. one random sequence of $\tilde{\epsilon}$ in Hamiltonian (\ref{RQKG}) and a single site initial excitation of the form $x_{500}(0)>0$, $x_i(0)=0$ for $1 \leq i \leq 1000$, $i \neq 500$, and $p_j(0)=0$ for $1 \leq j \leq 1000$. For the computation of the Lyapunov exponents, we apply the {\em tangent map method}\cite{Skokosetal2010,GES12} which is suitable for the evolution of deviation vectors in the tangent space of the orbit under study. Having thus access to the deviation vectors, we compute the Lyapunov exponents in descending order (i.e. $\lambda_1>\lambda_2\ldots >\lambda_{2N}$) following the so-called {\em standard method}\cite{Benettin1980a,Benettin1980b,S10}.

\subsection{Complex Statistics Shows Persisting Chaos in the Klein-Gordon Chain}\label{statistics}

In this section we shall view the values of one (or a linear combination) of coordinates of our solutions of Eq. (\ref{KG-EOM}), at discrete times $t_i,\;i=1,\ldots, M$, as realizations of $\mathcal{N}$ iid random variables $X^{(j)}(t_i)$, $j=1,\ldots,\mathcal{N}$. If these variables are random, according to the CLT, the distribution of their sums will yield a Gaussian pdf, whose mean and variance are those of the $X^{(j)}$'s. As described in the introduction, this is what happens in many dynamical systems in regions of strong (or uniform) chaos, where correlations decay exponentially and the system obeys Boltzmann-Gibbs statistics. In weakly chaotic regions however, pdfs of sums of orbit components {\em do not} rapidly converge to a Gaussian, but are well approximated, for long times, by the $q$-Gaussian (\ref{q_gaussian}) distribution characterized by $1\leq q<3$ ($q=1$ corresponding to a Gaussian pdf).

Let us start by examining in detail the statistical properties of the lattice as the initial excitation of a central particle starts to be transmitted to its neighboring sites. We focus on the time evolution of the $q$ entropic index for a class of observable functions that start with the central particle (i.e. $\eta_1=x_{500}$) and gradually take into account more and more sites symmetrically to the initially excited one, up to the whole extent of the lattice, i.e. $\eta_1=x_{500}$, $\eta_5=x_{498}+\ldots+x_{502}$, $\eta_9=x_{496}+\ldots+x_{504}$, $\eta_{19}=x_{491}+\ldots+x_{509}$, $\eta_{29}=x_{486}+\ldots+x_{514}$, $\eta_{39}=x_{481}+\ldots+x_{519}$, $\eta_{1000}=x_{1}+\ldots+x_{1000}$, where the subscript of $\eta$ denotes the number of particles considered in the computation of the observable function.

\begin{figure}[!ht]
\centering{
\includegraphics[scale=1]{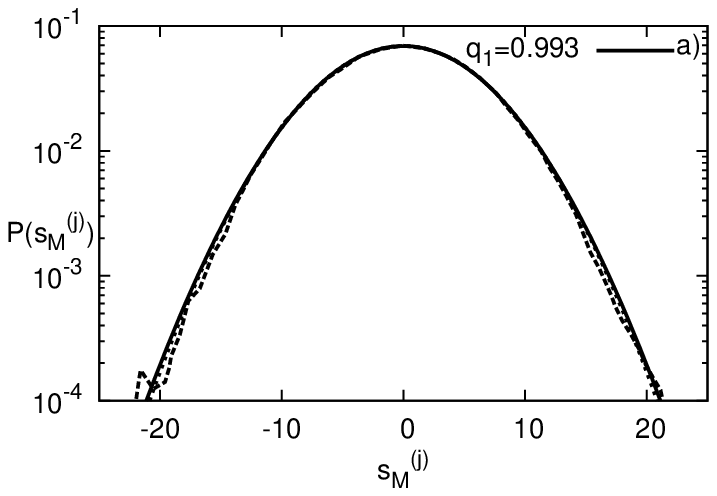}
\includegraphics[scale=1]{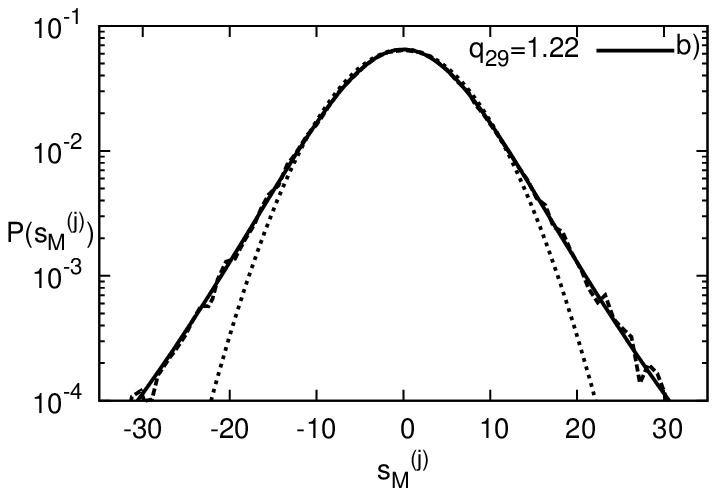}
\caption{Panel a): Plot of the numerically computed pdf (dashed curve) for the observable $\eta_1$ in the time interval $[0,10^8]$ with $q_1=0.993 \pm 0.009$ taken from fitting with a $q$-Gaussian distribution (\ref{q_gaussian}) in solid thick. Panel b): Similar plot of the numerically computed pdf (dashed) for the observable $\eta_{29}$ and a time interval $[0,10^8]$ with $q_{29}=1.22 \pm 0.01$. In both panels, $N=1000$ and $E=0.4$ that corresponds to the subdiffusive case. Note that the vertical axes are in logarithmic scale, while the dotted curve is the Gaussian pdf (i.e. $q=1$).}\label{distributions_N1000_E0.4_different_etas}}
\end{figure}

In Fig. \ref{distributions_N1000_E0.4_different_etas} we show two representative examples of numerical distributions with different $q$ entropic indices for the low energy subdiffusive case, i.e. $E=0.4$ (the initial value of $x_{500}(0)$ is adjusted so that $E=0.4$). In panel a) we plot the numerical distribution (dashed curve) for the observable $\eta_1$ computed in the time interval $[0,10^8]$ and find that it is well fitted by a $q$-Gaussian distribution (solid thick curve) with $q_1=0.993 \pm 0.009$. This is a case where the numerical distribution is indistinguishable from a Gaussian ($q=1$) plotted as a dotted curve. On the other hand, panel b) which is the same plot as a), for the observable $\eta_{29}$, reveals a clear $q$-Gaussian distribution (\ref{q_gaussian}), over nearly four decades on the vertical axis, with $q_{29}=1.22 \pm 0.01$. Note that we always plot the Gaussian pdf (i.e. $q=1$) as a dotted curve to guide the eye.

\begin{figure}[!ht]
\centering{
\includegraphics[scale=1]{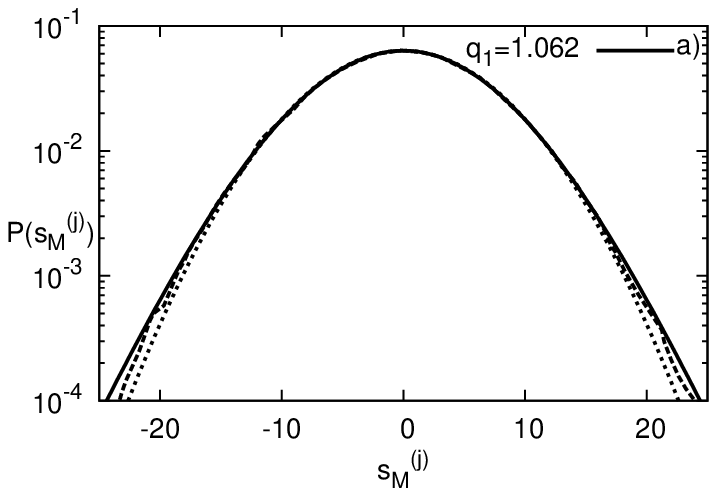}
\includegraphics[scale=1]{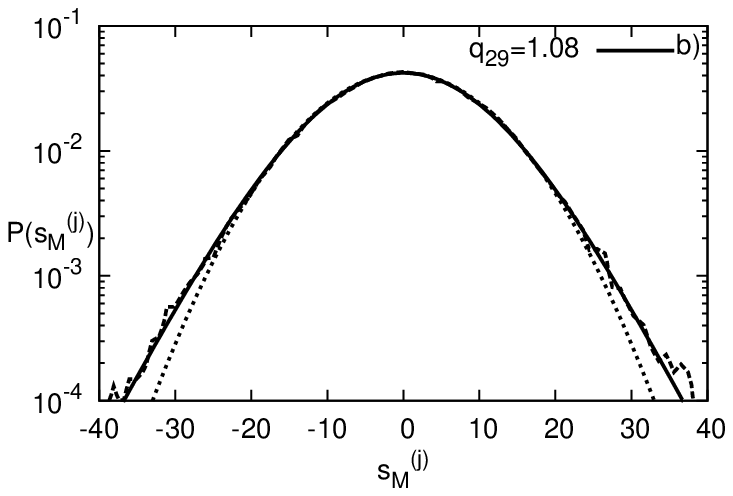}
\caption{Panel a): Plot of the numerically computed pdf (dashed curve) for $\eta_1$ in the time interval $[0,10^8]$ with $q_1=1.062 \pm 0.008$ fitted by a $q$-Gaussian distribution (solid thick curve). Panel b): Plot of the pdf (dashed curve) for $\eta_{29}$ and the time interval $[0,10^8]$ fitted by a $q$-Gaussian distribution (solid thick curve), with $q_{29}\approx1.08 \pm 0.01$. In both panels, we use $N=1000$ and $E=1.5$. Note that the vertical axes are in logarithmic scale, while the dotted curve is the Gaussian pdf (i.e. $q=1$).}\label{distributions_N1000_E1.5_different_etas}}
\end{figure}

Now, let us present the corresponding probability distributions for the self-trapping case $E=1.5$ in Fig. \ref{distributions_N1000_E1.5_different_etas} keeping everything else the same as in Fig. \ref{distributions_N1000_E0.4_different_etas} (similarly, the initial value of $x_{500}(0)$ is adjusted so that $E=1.5$). We see that in this case not only the entropic index $q_1$ but also $q_{29}$ is closer to the $q=1$ value of a Gaussian compared to that of Fig. \ref{distributions_N1000_E0.4_different_etas}b).

\begin{figure}[!ht]
\centering{
\includegraphics[scale=0.9]{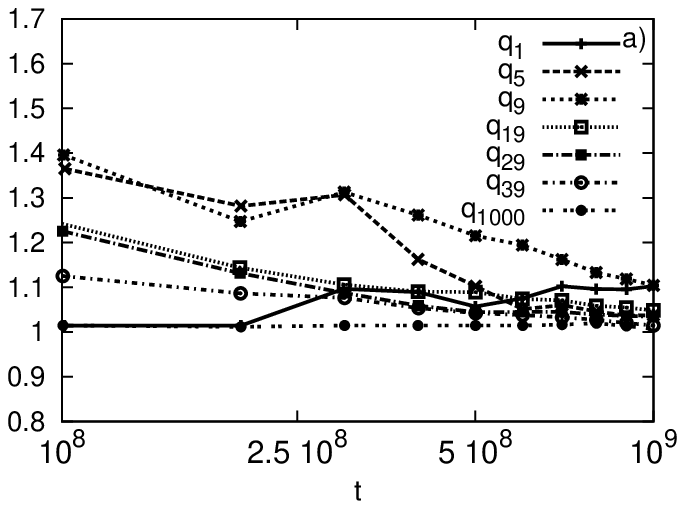}
\includegraphics[scale=0.9]{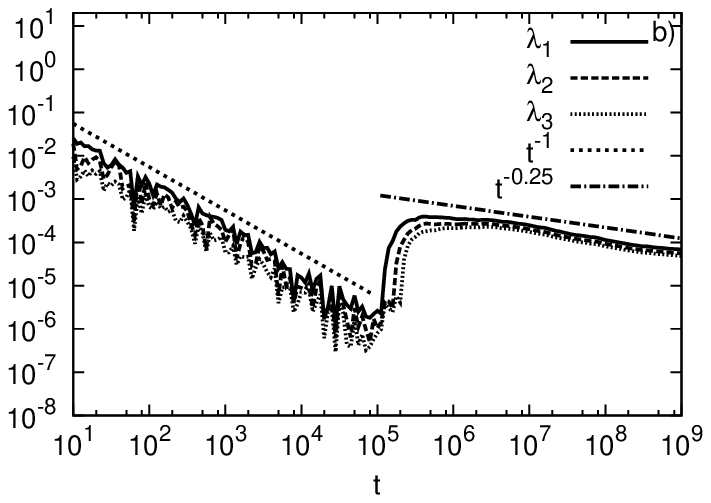}\\
\includegraphics[scale=0.9]{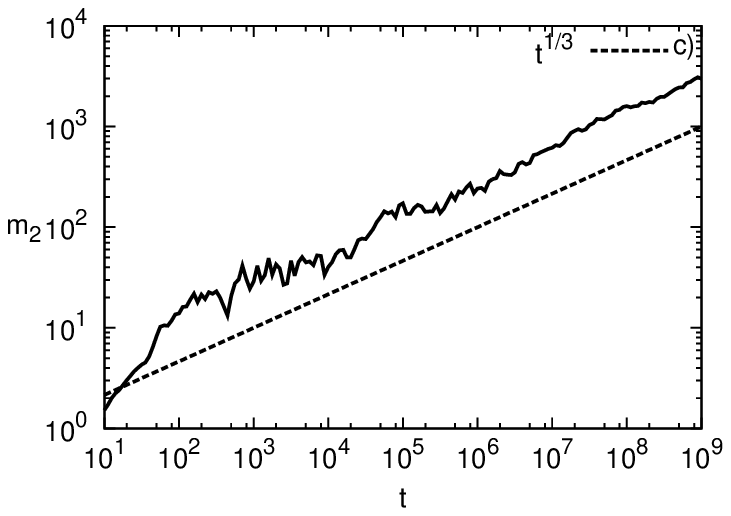}
\caption{Panel a): Plot of the time evolution of the $q$ entropic indices $q_1,\;q_5,\;q_9,\;q_{19},\;q_{29},\;q_{39}$ and $q_{1000}$ for $N=1000$ and $E=0.4$ that corresponds to the subdiffusive case. Panel b): Plot of the evolution of the corresponding three largest Lyapunov exponents $\lambda_1$, $\lambda_2$, $\lambda_3$, and of $t^{-1}$, $t^{-0.25}$ to guide the eye. Panel c): Plot of the corresponding second moment $m_2$ in time together with $t^{1/3}$ to guide the eye. Note that all horizontal axes are logarithmic.}\label{q_time_evolution_N1000_E0.4_different_etas}}
\end{figure}

Next, in panel a) of Fig. \ref{q_time_evolution_N1000_E0.4_different_etas}, we see that in the subdiffusive case ($E=0.4$), the central 5 to 29 particles initially perform a weakly chaotic motion depicted by the tendency of the corresponding $q_1$-$q_{29}$ entropic indices to attain values considerably higher than 1 (even though they later decay towards 1). On the other hand, if one includes more particles and studies $\eta_{39}$ for example, the motion is more chaotic since the corresponding entropic index now tends more quickly to 1 at $t=10^9$, while if we consider all particles (i.e. for $\eta_{1000}$) strong chaos becomes clearer as $q_{1000}$ tends to 1 even more rapidly.

These results suggest that the behavior of the central part of the lattice is more weakly chaotic, while the whole lattice behaves in a strongly chaotic way. This is also apparent in panel b) of Fig. \ref{q_time_evolution_N1000_E0.4_different_etas}, where the three largest Lyapunov exponents $\lambda_1,\lambda_2,\lambda_3$ initially show a tendency to decrease towards zero, however, after $t=10^5$ they suddenly jump to higher values and then decrease with a slope smaller than 1. Recently\cite{SGF13} it was found that for this case the maximum Lyapunov exponent $\lambda_1$ decreases as $\lambda_1 \propto t^{-0.25}$. This behavior is also seen in Fig. \ref{q_time_evolution_N1000_E0.4_different_etas}b). We note that we computed only a few Lyapunov exponents because the computation of many of them in a high dimensional system is a very hard computational task. From the results of Fig. \ref{q_time_evolution_N1000_E0.4_different_etas}b) it is evident that the evolution of these exponents is determined by the evolution of the maximum Lyapunov exponent $\lambda_1$, as all of them show similar behaviors. As we see from panel c) of Fig. \ref{q_time_evolution_N1000_E0.4_different_etas}, the expected behavior of the second moment, i.e. $m_2\propto t^{1/3}$, is well reproduced by our numerical results, which serves as additional evidence for our computational accuracy.

\begin{figure}[!ht]
\centering{
\includegraphics[scale=0.9]{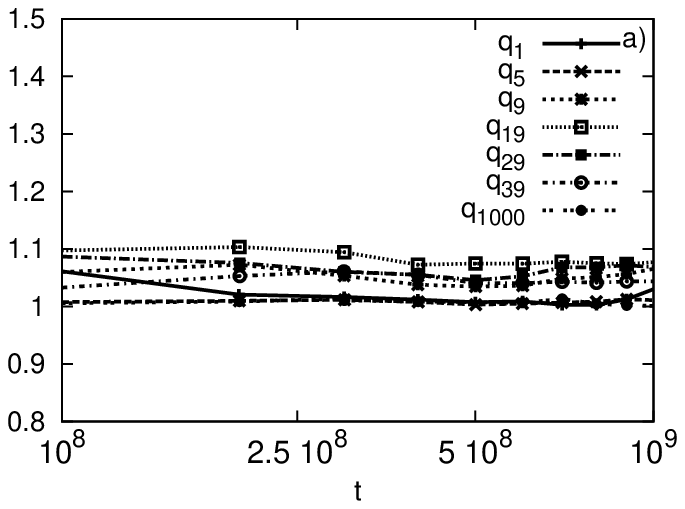}
\includegraphics[scale=0.9]{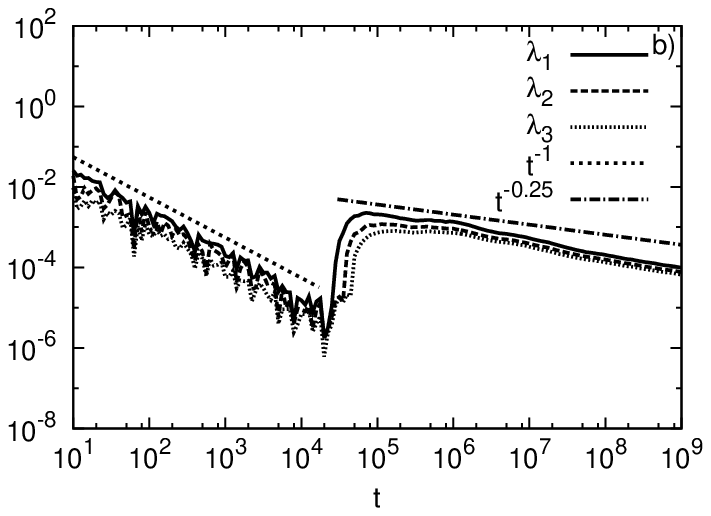}\\
\includegraphics[scale=0.9]{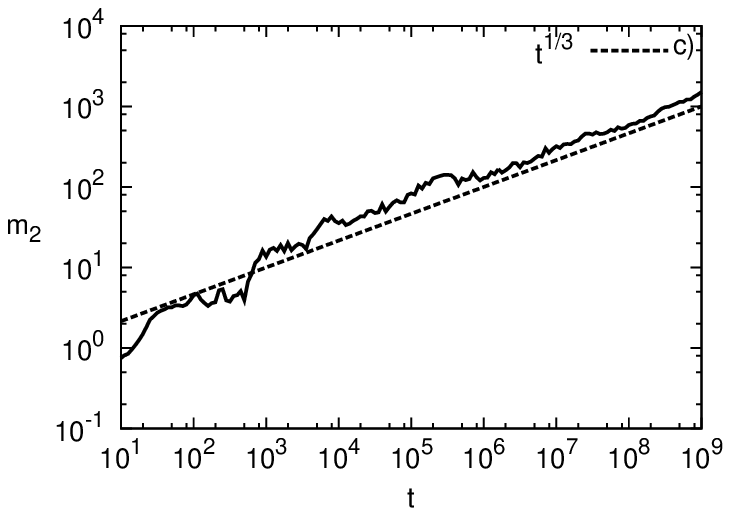}
\caption{Panel a): Plot of the time evolution of the $q$ entropic indices $q_1,\;q_5,\;q_9,\;q_{19},\;q_{29},\;q_{39}$ and $q_{1000}$ for $E=1.5$. Panel b): Time plot of the corresponding three largest Lyapunov exponents $\lambda_1$, $\lambda_2$, $\lambda_3$, and of $t^{-1}$, $t^{-0.25}$ to guide the eye. Panel c): Plot of the corresponding second moment $m_2$ in time together with $t^{1/3}$ to guide the eye. Note that all horizontal axes are logarithmic.}\label{q_time_evolution_N1000_E1.5_different_etas}}
\end{figure}

By contrast, in Fig. \ref{q_time_evolution_N1000_E1.5_different_etas} where the same study is repeated for the higher energy $E=1.5$ of the self-trapping case, the dynamics is somewhat different. Panel a) shows that all $q$ entropic indices of Fig. \ref{q_time_evolution_N1000_E0.4_different_etas} are now much closer to 1, even those of the central particles. Comparing the three largest Lyapunov exponents in the two cases, we see that at the higher energy of the self-trapping case (which corresponds to stronger nonlinearity) they jump to higher values at about $t=10^4$, i.e. one order of magnitude {\em earlier} than in the case of the lower energy of the subdiffusive case. After that point, the Lyapunov exponents start decaying to zero but a bit faster than the one ($\propto t^{-0.25}$) observed in the subdiffusive case of Fig. \ref{q_time_evolution_N1000_E0.4_different_etas}. We note again that $m_2$ grows in time as $t^{1/3}$ as it can be evidenced in panel c) of Fig. \ref{q_time_evolution_N1000_E1.5_different_etas}. 

The reader may wonder at this point how representative are the results presented in panel a) of Fig. \ref{q_time_evolution_N1000_E0.4_different_etas} and Fig. \ref{q_time_evolution_N1000_E1.5_different_etas}, since they are based on the computation of only two trajectories. For this reason, we wish to clarify that we actually studied 10 additional trajectories and averaged their results for $t=10^8$. The curves we obtained were not significantly different than what is shown in Fig. \ref{q_time_evolution_N1000_E0.4_different_etas}a) and Fig. \ref{q_time_evolution_N1000_E1.5_different_etas}a). There was a tendency of the $q$ values to fall towards 1 but since calculations for $t=10^9$ are very time consuming we decided to postpone a more detailed study for a future publication.

A comment on the numerical evaluation of $\lambda_1,\lambda_2,\lambda_3$ is useful here. In our computation we use three initially linearly independent deviation vectors, namely $(1,0,0,0, \ldots, 0)$, $(0,1,0,0, \ldots, 0)$ and $(0,0,1,0, \ldots, 0)$, which correspond to initial perturbations at the first three lattice sites. These oscillators remain practically unexcited for the duration of the numerical simulation, as long as the lattice size and the final integration time are chosen so that the wave packet does not reach the lattice boundaries. Nevertheless, due to the coupling between the oscillators, these perturbations eventually propagate throughout the whole lattice. The choice of the initial deviation vectors influences the initial phase of the Lyapunov exponents but not their asymptotic behavior, as any set of deviation vectors eventually leads to the same estimation. During this initial phase the estimated Lyapunov exponents behave as in the case of regular orbits, exhibiting a $\propto t^{-1}$ decrease. Different choices of the initial deviation vectors result in changing the duration of this phase, or even lead to its disappearance\cite{SGF13}. Nevertheless, using the {\em same} set of deviation vectors in both cases of Figs. \ref{q_time_evolution_N1000_E0.4_different_etas} and \ref{q_time_evolution_N1000_E1.5_different_etas} allows for a direct comparison between them.

\section{Conclusions}\label{conclusions}

In this paper we have studied the dynamics and statistics of diffusive motion in a 1-dimensional Klein-Gordon chain in the presence of disorder. Our statistical approach is based on the computation of sums of position coordinates, in the spirit of the Central Limit Theorem, approximating their pdfs by $q$-Gaussians, whose index $q>1$ is connected with weak chaos, while $q=1$ corresponds to strong chaos. 

In particular, we considered a disordered KG chain of $N=1000$ particles focusing on a low energy (subdiffusive) and on a higher energy (self-trapping) case and verified that subdiffusive spreading always occurs following specific power-laws with exponents smaller than 1 as pointed out in the literature. Subsequently, integrating the equations of motion for times as long as $10^9$ and computing the corresponding pdfs, we found evidence that the dynamics does {\em not} relax onto a quasi-periodic KAM torus, as it has been conjectured, but continues to spread chaotically along the chain for arbitrarily long times.

One might argue that, since the strength of the nonlinearity diminishes during spreading, it is somewhat counter-intuitive that the motion should become more chaotic with $q\rightarrow1$ as time grows. We conjecture however, that this may be due to the fact that, as diffusion progresses and more particles become activated, the effective \textit{dimensionality} of the phase space becomes higher and hence the orbits have to wander over wider chaotic domains, while stickiness on the boundaries of multidimensional islands becomes less likely to happen.

Recently, we carried out a preliminary study of systems of $N$ coupled 2-dimensional symplectic maps that may be regarded as simple examples of Hamiltonian particle chains. Indeed, it would be highly desirable to find such examples, as they would permit a much more detailed investigation of diffusive phenomena, owing to their great computational advantage over systems of ordinary differential equations. However, even though we do find examples of such systems that display subdiffusive motion with pdfs of the $q$-Gaussian type, there are important differences that, at present, preclude their use as alternative models for the type of diffusive dynamics we have studied in this paper. This is an interesting topic which we plan to address in a future publication.

\begin{acknowledgments}
We would like to thank the referees for their constructive criticism and many useful suggestions that helped us considerably improve the presentation of our results. One of the authors (T. B.) gratefully acknowledges the hospitality of the New Zealand Institute of Advanced Study during the period of February 20 - April 15, 2013, when some of this work was carried out. He is thankful for many useful conversations he had during his stay with Professor Sergej Flach on topics related to the content of the present paper. This research has been co-financed by the European Union (European Social Fund - ESF) and Greek national funds through the Operational Program ``Education and Lifelong Learning'' of the National Strategic Reference Framework (NSRF) - Research Funding Program: THALES - Investing in knowledge society through the European Social Fund. Ch. S. was also supported by the Research Committees of the University of Cape Town (Start-Up Grant, Fund No 459221) and of the Aristotle University of Thessaloniki (Prog. No 89317). Computer simulations were performed in the facilities offered by the HPCS Lab of the Technological Educational Institute of Western Greece.
\end{acknowledgments}


\providecommand{\noopsort}[1]{}\providecommand{\singleletter}[1]{#1}%

\end{document}